\documentclass[11pt,eps]{article}

\usepackage{graphicx}
\usepackage{epstopdf}
\begin{document}

\begin{center}
{BR-EP-33-05-01-06 revised 10-03-06 and 10-26-06 1, submitted for publication}
\vskip0.30cm

{\large {\bf Fundamental Neutrino Experiments}}

{\bf Ruggero Maria  Santilli\\ Institute for  Basic Research, 
Box  1577, Palm Harbor,  FL 34682,  U.S.A.\\
  ibr@gte.net, http://www.i-b-r.org}
\end{center}

\begin{abstract}
We review fundamental open problems in neutrino physics and propose two basic experiments for their possible resolution.
\end{abstract}

\noindent Following Rutherford's prediction of the synthesis of the neutron inside stars from protons and electrons [1a], Chadwick confirmation [1b], Pauli's objection for the lack of spin ${1\over 2}$, Fermi's hypothesis of the neutrino (meaning "little neutron" in Italian) [1d], and other advances, the neutrino hypothesis is today part of the {\it standard model} (see, e.g., [1e].

Despite clearly historical results, the neutrino physics has remained afflicted by  fundamental, unresolved, theoretical and experimental problems, such as:

1) Neutrino physics is based on an excessive number of individually unverifiable assumptions. In fact, the original hypothesis of one massless neutrino, was replaced by the sequential 
hypotheses that: there exist three different neutrinos and their antiparticles; neutrinos have masses;   neutrino  masses are different; neutrinos  ``oscillate''; with additional hypotheses expected due to the known insufficiencies  of the preceding ones.

2) According to the standard model, said various neutrinos  can traverse very large hyperdense media such (as entire stars) without any collision while being {\it massive particles carrying energy in our spacetime.} This view is beyond scientific plausibility.

3) The familiar reaction $p^+ +  e^-\rightarrow n + \nu$ violates the conservation of the
energy unless the proton and the electron have a kinetic 
energy of at least $0.78 MeV$, in which case there is no energy left for the neutrino. In fact,  the neutron rest energy ($939.56 MeV$) is $0.78 \,MeV$ {\it bigger} than the sum of
the rest energies of the proton and the electron ($938.78  MeV$).

4) Calculations on the ``bell shaped''  form of the energy of the electron in nuclear beta decays $N(A, Z)\rightarrow  N(A, Z+1) + e^- + \bar \nu$ show that no energy  appears to be left for the neutrino, provided hat nuclei  are represented in their actual extended size. In fact,    
the Coulomb interaction between an {\it extended} nucleus and the emitted electron varies
with the direction of the beta emission, with maximal (minimal) value for radial (tangential) emissions,  the ``missing energy'' being apparently absorbed by the nucleus.

5) Neutrino experiments are perhaps more controversial than theoretical studies because: the number of events used as "experimental evidence" for the existence of neutrinos  is excessively small over an extremely large number of events, thus preventing acceptance by the physics community at large; experimental data are elaborated with a theory crucially dependent on the existence of the neutrinos, in which case the "experimental results" are expected to depend on the theoretical assumptions; the theory contains an excessive number of parameters (such as the different neutrinos masses and others) essentially capable to achiever any desired fit; some of the recent "neutrino detectors" contain radioactive isotopes that could themselves trigger the very few selected events; and other reasons.

For additional insufficiency studies, one may consult Bagge [2a] and Franklin [2b] for alternative theories without neutrinos; Wilhelm [2c] for additional problematic aspects; M\"ossbauer [2d] for problems in neutrino oscillations; Fanchi [2e] for apparent biases in "neutrino experiments";  and references quoted therein.

A main point attempted to convey in this note is that the above issues can be best solved via theoretical and experimental studies on the neutron synthesis and decay. The historical difficulty is that, while neutrons are unquestionably synthesized from protons and electrons inside stars, such a synthesis cannot be consistently treated via quantum mechanics for numerous reasons, such as: 1) The Schr\"odinger equation becomes physically inconsistent for the "positive" binding-like energies needed for the reaction $p^+ + e^-\rightarrow n + \nu$, as the reader is encouraged to verify by trying to solve any quantum bound state in which the conventional negative binding energy is turned into a positive value; 2) Quantum mechanics cannot represent the spin of the neutron as per historical objections [1c.1d]; and, additionally, 3) Quantum mechanics cannot provide a representation of the meanlife, charge radius, and anomalous magnetic moment of the neutron.

In view of these limitations, comprehensive studies have been conducted  for the construction of a broader nonunitary realization of the (abstract) axioms of quantum mechanics (qm) under the name of {\it hadronic mechanics} (hm) [3a-3c]. The main  idea is that quantum mechanics is assumed to be exactly valid for all mutual distances of particles permitting their credible approximation as being point-like, while at smaller  mutual distances (e.g., of about $1 F$) nonunitary corrections are admitted to represent contact, nonlinear, nonlocal and non-Hamiltonian effects expected in deep mutual  penetrations of particles. 

Following applications and verifications in various fields [{\it loc. cit.}], the new mechanics did achieve an exact and (time) invariant nonrelativistic [3d] and relativistic [3e] representation of {\it all} characteristics of the neutron synthesized from a hydrogen atom (see [3f,3h] for reviews). Physical consistency of Schr\"odinger's equation is regained because nonunitary transforms  cause a new renormalization (called {\it mutations}) of the rest energy and other particle features under which the binding energy returns to be negative. A nonunitary transform  $UU^\dag \not = I, Lim_{r>> 1 F} (UU^\dag) = I$ of the conventional equations for the hydrogen atom then permitted the exact and invariant representation, with one single structural equation, of  the rest energy, charge radius, meanlife, charge and parity of the neutron.

The representation of the neutron spin turned out to be easier than expected. To avoid large repulsive forced as occurring in gears, the electron is constrained to penetrate within the proton in singlet coupling and to orbit with an angular momentum equal to the proton spin (otherwise the electron has to rotate against the hyperdense medium inside the proton) resulting in a null total angular momentum. Consequently, in this model the spin of the neutron coincides with the proton spin. The exact representation of the neutron magnetic moment follows from the contribution due to the (generally ignored) orbital motion of the electron within the proton.  It should be noted that an angular momentum with value ${1\over 2}$ is anathema for quantum mechanics, but fully acceptable for the covering hadronic mechanics precisely in view of its nonunitary structure [3f,3h]. 

 The model is generally indicated with the symbol $n = (\hat p^+, \hat e^-)_{hm}$, where the "hats" denote particles "mutated" (or deformed) by their deep mutual penetration or, more technically, characterized by irreducible representations of the nonunitary covering of the Poincar\'e symmetry known as the {\it Poincar\'e-Santilli isosymmetry} (see [3a] for details and large bibliography). Note that the conventional Poincar\'e symmetry cannot be credibly claimed to be absolutely "exact" for the structure of the neutron as well as hadrons at large due to the lack of a planetary structure [3a].
 
 Intriguingly, the model admits only {\it one} energy level, that of the neutron. Consequently, the "excited states" of the neutron are given by the conventional energy states of the hydrogen atom, since at  distances bigger than $1 F$ quantum mechanics is regained exactly and uniquely. The extension of the model to the remaining hadrons and its compatibility with the standard model  are discussed in Ref. [3f].

To continue these studies, Santilli [3f] has recently submitted the  hypothesis of the {\it etherino} (meaning "little ether" in Italian) with symbol "a" (from the Latin aether), representing an "entity" with mass and charge $0$, spin ${1\over 2}$, and energy $0.78 MeV$ and new reaction $p^+ + a + e^-\rightarrow n$. The missing quantities carried by the etherino are assumed to originate either from the interior of stars or from the ether conceived as a universal medium with very high energy density. The latter possibility was submitted to allow quantitative studies of the old hypothesis of continuous creation of matter in the universe since its most plausible realization could occur precisely in the synthesis of neutrons inside stars [3f]. Note that the etherino represents the {\it transfer of quantities} from the medium to the neutron and { it {\it does not} represent a particle as conventionally understood.

The need {\it at the quantum level} of a new entity for the neutron synthesis originates from several reasons [3f], including: the impossibility of using an antineutrino for delivering the missing quantities according to the reaction $p^+ + \bar \nu + e^-\rightarrow n$ due to impossibility of any realistic scattering of $\bar \nu$ with $p^+$ and/or $e^-$; the impossibility of assuming that the proton and the electrons have the missing (relative) energy of $0.78 MeV$ because at that value their cross section is extremely small (of the order of $10^{-20} barns$); and other reasons [3a].

It should be stressed that {\it at the level of hadronic mechanics there is no need for the etherino hypothesis} due to the equivalence $(p^+, a, e^-)_{qm} \approx (\hat p^+, \hat e^-)_{hm}$. In fact, the nonunitary covering of the Hilbert space (called {\it iso-Hilbert space} [3a]) is a direct representation of the missing quantities in the neutron synthesis [3a,3f]. One should keep in mind  that the l.h.s of the preceding expression is not solvable, while the r.h.s is exactly solvable.

Note that {\it the (quantum) etherino hypothesis is not necessarily in conflict with the (quantum) neutrino hypothesis,} because the former deals with the neutron {\it synthesis} while the latter deals with the neutron {\it decay}, and we can indeed have the sequence $p^+ + a + e^-\rightarrow n\rightarrow p^+ +  \bar \nu + e^- $. However, it should be indicated that the neutron decay according to hadronic mechanics does not  mandate the emission of a neutrino, because the  Poincar\'e-Santilli isosymmetry for non-planetary structures allows the conversion of rotational into linear motions (e.g., via constraints) as for the sling shot, with sequence $(\hat p^+, \hat e^-)_{hm} = n\rightarrow p^+ + e^-$. Additionally, we may have the sequence $p^+ + a + e^- \rightarrow n \rightarrow p^+ +  a + e^-$ where the final $a$ represents the return of the originally missing quantities to the medium. In con short, despite studies on the issue for decades, this author knows of no truly conclusive theoretical argument either in favor or against the neutrino hypothesis.
  


\begin{figure}[htbp] 
   \centering
   \includegraphics[width=2in]{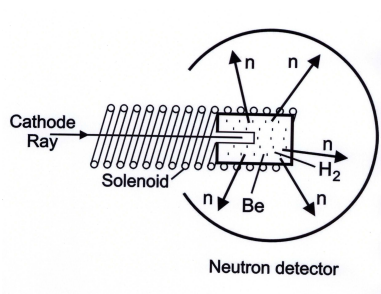} 
  \caption{{\it The proposed laboratory synthesis of the neutron from protons and electrons. }}
\end{figure}


 It should be also indicated that the possible lack of existence of the neutrino is not necessarily in conflict with neutrino experiments, since the new vistas  could merely require their reinterpretation, e.g., via impulses $a$ triggered by the neutron decay 
 and propagating through the ether, $n\rightarrow (p^+ + e^-)_{spacetime} + a_{ether}$, thus eliminating the assumption that massive particles  can cross entire stars without any collision. This possibility could initiate research for new  interstellar communications because the propagation of the "impulse" $a$ through the ether can only be
 longitudinal, thus having  a multiple of the speed of conventional (transversal) electromagnetic waves. 

The experimental community cannot continue to leave neutrino physics fundamentally unresolved without risking a serious condemnation by posterity. In the hope of stimulating the termination of the current condition, in this note we propose the following experiments that could establish the existence of the neutrino in a final form or support broader vistas.

{\bf Proposed experiment on the neutron synthesis.}
There is no doubt that neutrino physics will remain fundamentally unresolved until we have resolutory experiments on the neutron synthesis. The  first laboratory synthesis of the neutron from protons and electrons  known to this author was conducted  in Brazil by the Italian priest-physicist Don Carlo Borghi and his associates [3g] (for a review, see [3h]). The  tests were apparently successful, although they do  not allow the measurement of the {\it  energy} needed for the synthesis.

The latter information can be obtained in a variety of ways. That recommended in this note   consists in sending a coherent electron beam  against a beryllium mass saturated with hydrogen and kept at low temperature (so  that the protons of the hydrogen atoms can be approximately
considered to be at rest). A condition for credibility  is that said protons and electrons be
polarized to have antiparallel spins due to large repulsions in triplet couplings indicated earlier. Since the proton and the electron have opposite charges, said polarization can be achieved with the same  magnetic field as illustrated in Fig. 1.



\begin{figure}[htbp] 
   \centering
   \includegraphics[width=2in]{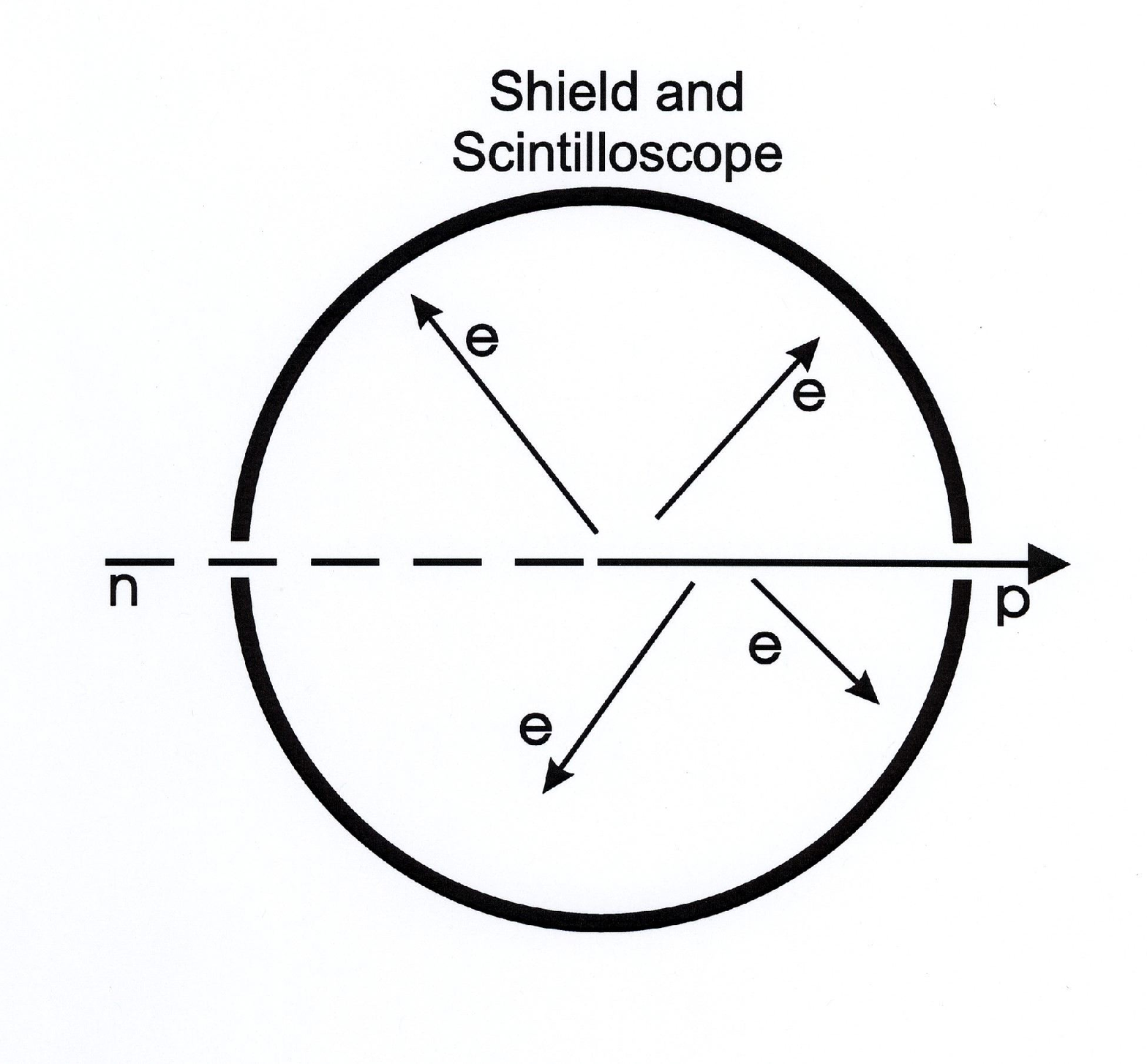} 
  \centering   \caption{{\it The proposed measurement of the electron energy  in neutron decays. }}\end{figure}


Neutrons that can possibly be synthesized in this way will escape from the beryllium mass and can be detected  with standard means. The detection of neutron produced with electron kinetic  energies systematically  in excess of $0.78 MeV$ would provide final confirmation of the neutrino hypothesis. The systematic detection of neutrons  synthesized either at the threshold energy of $0.78 MeV$ or less would support alternative hypotheses [2,3f], and render plausible the old hypothesis of continuous  creation of matter in the universe via the neutron synthesis. Note that the latter possibility {\it would not} necessarily deny the existence of the neutrino because experimental results on the neutron {\it synthesis} cannot be credibly claimed to apply necessarily for the neutron {\it decay}.

{\bf Proposed experiment on the neutron decay.} It is also clear that neutrino physics will remain  basically unresolved until we have new experimental data on the spontaneous neutron decay. This information can be reached in numerous ways. That recommended in this
note is to conduct systematic measurements of the energy of the electron emitted in the decay
of a coherent beam of {\it low energy} (e.g., thermal) neutrons as depicted in Fig. 2.  The systematic detection of electron energies less than  $0.78 MeV$ (plus the neutron energy) {\it would not} necessarily confirm the existence of the neutrino due to the alternative decays $n\rightarrow p^+ + \bar \nu + e^-$ or $n\rightarrow p^+ + \bar a + e^-$. The detection of electron energies systematically given by  $0.78 MeV$ (plus the neutron energy) would support alternative vistas [2,3f].   

Note that the conduction of  the proposed test with ``high energy" neutrons would not be resolutory  because the {\it variation} of the electron energy expected to be absorbed by the neutrino would be excessively smaller than   the electron energy.   The conduction of the test via nuclear beta decays is also not recommendable due to the indicated expected dependence of the electron energy from the direction of beta emission, which dependence is ignorable for the case of decay of individual neutrons. The author has been unable to identify in the literature any conduction of the proposed test since all available experiments refer to {\it nuclear} beta decays rather than that of individual {\it neutrons}.

\end{document}